# New Approach to Quantum Error Correction

Ri Qu, Bing-jian Shang, Yan-ru Bao and Yi-ping Ma

*School of Computer Science and Technology, Tianjin University, Tianjin, 300072, China and Tianjin Key Laboratory of Cognitive Computing and Application, Tianjin, 300072, China*

Operator quantum error correction provides a unified framework for the known techniques of quantum error correction such as the standard error correction model, the method of decoherence-free subspaces, and the noiseless subsystem method. We first show an example of a new quantum error correction scheme which can not be described by operation quantum error correction. Then we introduce a different notion of noiseless subsystems according to the example. Base on this notion, we present a more unified approach which incorporates operator quantum error correction as a special case. Moreover, we also give a sufficient and necessary condition of quantum error correction using this approach. We show that this approach provides more recovery operations than operator quantum error correction, which possibly leads to simpler decoding procedures.

## 1. Introduction

To develop quantum information processing technologies, it is necessary to protect quantum information against undesirable noise. Since 1990s, there has been considerable progress towards this goal. Fault-tolerant quantum computing theory [1-6] is built on standard model of *quantum error correction* (QEC) [1, 2, 4, 7]. In this QEC model, encoded quantum states are restricted to a code subspace $C$ of the whole system's Hilbert space $H = C \oplus C^\perp$. Several passive QEC methods, including *decoherence-free subspaces* [8-10] and *noiseless subsystems* [11-13], have also been presented. Refs. [14, 15] introduce the so-called *operator quantum error correction* (OQEC) that can describe all aforementioned QEC methods. OQEC relies on a generalized notion of a noiseless subsystem. In this notion, information is encoded in a subsystem $A$ of the code subspace $C = H^A \otimes H^B$. Exactly speaking, given a quantum channel $\varepsilon$, we can say that $A$ is a noiseless subsystem ($B$ is called a noisy subsystem) if

$$\forall \rho^B \forall \rho^A \exists \sigma^B : \varepsilon\left(\rho^A \otimes \rho^B\right) = \rho^A \otimes \sigma^B. \qquad (1)$$

where we write $\rho^A$ for density operators on $H^A$, similarly for $\rho^B$ and $\sigma^B$. In this letter, the above noiseless subsystem is called a "normal" noiseless one.

One might ask whether there is some new method which can not be described by OQEC. Our answer is "Yes". To see this, we discuss the following example.

**Example.** Consider a quantum channel $\varepsilon = \{E_0, E_1, E_2\}$ obtained as follows. Fix $0 < \gamma < 1$,

and with respect to the computational basis $\{|0\rangle, |1\rangle\}$ let $F_0 = \begin{bmatrix} 1 & 0 \\ 0 & \sqrt{1-\gamma} \end{bmatrix}$, $F_1 = \begin{bmatrix} 0 & \sqrt{\gamma} \\ 0 & 0 \end{bmatrix}$,

and $I_2 = \begin{bmatrix} 1 & 0 \\ 0 & 1 \end{bmatrix}$. Then define $E_0 = F_0 \otimes |0\rangle\langle 1|$, $E_1 = F_1 \otimes |0\rangle\langle 1|$, and $E_2 = I_2 \otimes |1\rangle\langle 0|$.

Clearly, $\sum_{a=0}^{2} E_a^\dagger E_a = I_2 \otimes I_2$.

Decompose a Hilbert space $\mathbb{C}^4 = H^A \otimes H^B$ with the computational basis, so that $H^A = \mathbb{C}^2$ and $H^B = \mathbb{C}^2$. For all $\rho^A$, we have

$$\varepsilon(\rho^A \otimes |1\rangle\langle 1|) = \left( \sum_{a=0}^{1} F_a \rho^A F_a^\dagger \right) \otimes |0\rangle\langle 0|, \tag{2}$$

$$\text{and } \varepsilon(\rho^A \otimes |0\rangle\langle 0|) = \rho^A \otimes |1\rangle\langle 1|. \tag{3}$$

It is known that the subsystem $A$ is not normal noiseless by (2). However, (3) implies that information encoded in the subsystem $A$ is protected against noise, which motivates us to define a different noiseless subsystem: information is encoded in the subsystem $A$ while the state of the subsystem $B$ is $|0\rangle$. Note that the different noiseless system is "one-time".

According to the above example, we consider the following condition: For a fixed density operator $\rho^B$,

$$\forall \rho^A \exists \sigma^B : \varepsilon(\rho^A \otimes \rho^B) = \rho^A \otimes \sigma^B. \tag{4}$$

And we can directly obtain that $(1) \Rightarrow (4)$. Conversely, the implication $(4) \Rightarrow (1)$ does not hold in general. Thus the condition (1) is stricter than the one (4). Thus, in this letter we will define more generalized noiseless subsystems, called amplicate noiseless subsystems, according to the condition (4). Furthermore, we will introduce a more unified approach that incorporates OQEC as a special case by means of amplicate noiseless subsystems. Moreover, we find that our approach is also guessed in terms of private quantum subsystems in Ref. [16].

We now describe our notations and nomenclatures. Let $\dim(H)$ be the dimension of a Hilbert space $H$. The set of *linear operators* on $H$ is denoted by $\mathcal{B}(H)$. In particular, the identity operator on $H$ is denoted $I^H$. It is known that a *density operator* $\rho$ can be written into $\sum_k p_k |\beta_k\rangle\langle\beta_k|$ where the eignstates $|\beta_k\rangle$ are corresponding to positive eignvalues $p_k$ and $\sum_k p_k = 1$. Thus we denote $\rho = \sum_k p_k |\beta_k\rangle\langle\beta_k|$ by the set $\{p_k, |\beta_k\rangle\}$. The *support*

$\sup(\rho)$ of $\rho$ is defined as the space spanned by $\{|\beta_k\rangle\}$. For convenience, we use $\rho$ and $\sigma$ for density operators. We refer to a *quantum channel (operation)* as a completely positive and trace-preserving convex linear map $\varepsilon: \mathcal{B}(H) \to \mathcal{B}(H)$. It is known that a channel $\varepsilon$ can be written into the *Choi-Kraus operation-sum form* $\varepsilon(\rho) = \sum_a E_a \rho E_a^\dagger$ for some operators $E_a \in \mathcal{B}(H)$ satisfying $\sum_a E_a^\dagger E_a = I^H$. Moreover, we denote the channel $\varepsilon$ by the error set $\{E_a\}$.

## 2. Ampliate Noiseless Subsystems

Let us decompose a Hilbert space $H = C \oplus C^\perp$ where $C = H^A \otimes H^B$. To show the essential difference between (1) and (4), we will discuss the conditions which are equivalent with (4) as follows. Note that the condition (ii) in the following lemma is corresponding to (4).

**Lemma 1.** Given a channel $\varepsilon$ and a density operator $\rho = \{p_k, |\beta_k\rangle\} \in \mathcal{B}(H^B)$, then the following three conditions are equivalent.

(i) $\exists \sigma^B \forall \rho^A : \varepsilon(\rho^A \otimes \rho) = \rho^A \otimes \sigma^B$;

(ii) $\forall \rho^A \exists \sigma^B : \varepsilon(\rho^A \otimes \rho) = \rho^A \otimes \sigma^B$;

(iii) $\forall k \exists \sigma_{\beta_k}^B \forall \rho^A : \varepsilon(\rho^A \otimes |\beta_k\rangle\langle\beta_k|) = \rho^A \otimes \sigma_{\beta_k}^B$.

*Proof.* The implication (i)$\Rightarrow$(ii) is trivial. It is clear that (iii)$\Rightarrow$(i) according to the linearity of $\varepsilon$. In the following, we prove that (ii)$\Rightarrow$(iii). The condition (ii) implies that for all pure state $|\psi\rangle \in H^A$,

$$\sum_k p_k \varepsilon(|\psi\rangle\langle\psi| \otimes |\beta_k\rangle\langle\beta_k|) = |\psi\rangle\langle\psi| \otimes \sigma^B \tag{5}$$

for some $\sigma^B \in \mathcal{B}(H^B)$. Then for all $k$ we can identify the form of $\varepsilon(|\psi\rangle\langle\psi| \otimes |\beta_k\rangle\langle\beta_k|)$ as follows. Suppose that $\{|\psi_j\rangle\}_{j=1,\ldots,\dim(H^A)}$ is a normal orthogonal basis of $H^A$ and $|\psi_1\rangle = |\psi\rangle$. Let $\lambda$ be a positive eignvalue of $\sigma^B$ and $|\phi\rangle$ be one of its eignstates. According to (5), we can obtain

$$\langle\psi_j|\langle\phi|\sum_k p_k \varepsilon(|\psi\rangle\langle\psi| \otimes |\beta_k\rangle\langle\beta_k|)|\psi_j\rangle|\phi\rangle = \begin{cases} \lambda & j=1 \\ 0 & others \end{cases}. \tag{6}$$

Since $\varepsilon$ is completely positive and trace-preserving, for all $k$ there is a density operator $\sigma_{\psi,\beta_k}^B$

such that

$$\varepsilon(|\psi\rangle\langle\psi|\otimes|\beta_k\rangle\langle\beta_k|) = |\psi\rangle\langle\psi|\otimes\sigma^B_{\psi,\beta_k}. \qquad (7)$$

According to a linearity argument, $\sigma^B_{\psi,\beta_k}$ does not depend on $|\psi\rangle$, that is, $\sigma^B_{\beta_k} \equiv \sigma^B_{\psi,\beta_k}$. Since $|\psi\rangle$ is chosen arbitrarily, $\sigma^B_{\beta_k}$ is invariant for all $\rho^A$. ∎

From the above lemma, (4) can be described as the following condition: For two fixed density operators $\rho_1, \rho_2 \in \mathcal{B}(H^B)$,

$$\forall \rho^A : \varepsilon(\rho^A \otimes \rho_1) = \rho^A \otimes \rho_2. \qquad (4')$$

Now we discuss the relationship between (4') and (1). According to the supports of $\rho_1$ and $\rho_2$, there are three cases as follows. For convenience, let $H^{B_1} = \sup(\rho_1)$. a) $H^{B_1} = H^B$. By (iii) in the lemma 1 and the linearity of $\varepsilon$, it is known that (4') is equivalent to

$$\forall \rho^A \exists \sigma^B : \varepsilon\left(\rho^A \otimes \frac{1}{\dim(H^B)} I^{H^B}\right) = \rho^A \otimes \sigma^B. \qquad (8)$$

According to the lemma 2 in Ref. [14], it is clear that (8) is equivalent to (1), which implies (4') and (1) are equivalent, that is, $A$ is a normal noiseless subsystem if the condition (4') holds; b) $\dim(H^{B_1}) < \dim(H^B)$ and $\sup(\rho_2)$ is a subspace of $H^{B_1}$. It is clear that $\rho_1, \rho_2 \in \mathcal{B}(H^{B_1})$. We give a revised decomposition $H = C' \oplus C'^\perp$ where $C' = H^A \otimes H^{B_1}$. Then this case is the same as a). Thus we obtain that (4') is equivalent to

$$\forall \rho^{B_1} \forall \rho^A \exists \sigma^{B_1} : \varepsilon(\rho^A \otimes \rho^{B_1}) = \rho^A \otimes \sigma^{B_1}. \qquad (9)$$

Thus the subsystem $A$ is still a normal noiseless one if the condition (4') holds; c) $\dim(H^{B_1}) < \dim(H^B)$ and $H^{B_1}$ does not include $\sup(\rho_2)$. It is known that $\rho_1 \in \mathcal{B}(H^{B_1})$ and $\rho_2 \notin \mathcal{B}(H^{B_1})$. According to (iii) in the lemma 1 and the linearity of $\varepsilon$, it is known that (4') is equivalent to

$$\forall \rho^A \exists \sigma^B : \varepsilon\left(\rho^A \otimes \frac{1}{\dim(H^{B_1})} I^{H^{B_1}}\right) = \rho^A \otimes \sigma^B. \qquad (10)$$

Clearly, (8) and (10) are not equivalent, which implies that (4')$\Rightarrow$(1) does not hold. Thus $A$ is not a normal noiseless subsystem if the condition (4) holds. But the quantum information encoded in $\rho^A$ is indeed immune to noise in the channel $\varepsilon$. And the state $\rho_1$ in the "small" subsystem $B_1$ is mapped into some state in another "big" subsystem $B$ under $\varepsilon$. This means that the space of the noisy subsystem is "amplified" from $H^{B_1}$ to $H^B$, that is, the decomposition of $H$ is transformed from $H = C' \oplus C'^\perp$ ($C' = H^A \otimes H^{B_1}$) to $H = C \oplus C^\perp$ ($C = H^A \otimes H^B$). However, the decomposition of $H$ is fixed for normal noiseless subsystems. Thus this case can be regarded as the synthesis of the above cases a) and b). Clearly, this case can not be described by the known QEC approaches.

To define a more general notion of a noiseless subsystem according to (4'), we give the following lemma. Note that (4') is corresponding to (i) in the following lemma.

**Lemma 2.** Decompose a Hilbert space $H = C \oplus C^\perp$ where $C = H^A \otimes H^B$. Let $\varepsilon$ be a channel.and $H^{B_1}$ be a subspace of $H^B$. Then the following three conditions are equivalent.

(i) There exist two density operators $\rho_1 \in \mathcal{B}(H^{B_1})$ and $\rho_2 \in \mathcal{B}(H^B)$ such that $H^{B_1} = \sup(\rho_1)$ and $\forall \rho^A : \varepsilon(\rho^A \otimes \rho_1) = \rho^A \otimes \rho_2$;

(ii) $\forall \rho^{B_1} \forall \rho^A \exists \sigma^B : \varepsilon(\rho^A \otimes \rho^{B_1}) = \rho^A \otimes \sigma^B$;

(iii) There exists a density operator $\rho \in \mathcal{B}(H^B)$,

$$\forall \rho^A : \varepsilon\left(\rho^A \otimes \frac{1}{\dim(H^{B_1})} I^{H^{B_1}}\right) = \rho^A \otimes \rho.$$

*Proof.* By the lemma 1, (ii)$\Rightarrow$(i) is trivial. Firstly, we prove (i)$\Rightarrow$(iii). From (i), we have $\forall \rho^A : \varepsilon(\rho^A \otimes \rho_1) = \rho^A \otimes \rho_2$ for $\rho_1 = \{p_k, |\beta_k\rangle\}_{k=1,\ldots,\dim(H^{B_1})}$. Then we can obtain that $\forall k \exists \sigma^B_{\beta_k} \forall \rho^A : \varepsilon(\rho^A \otimes |\beta_k\rangle\langle\beta_k|) = \rho^A \otimes \sigma^B_{\beta_k}$ according to the lemma 1. By the linearity of $\varepsilon$, we have $\forall \rho^A : \varepsilon\left(\rho^A \otimes \frac{1}{\dim(H^{B_1})} I^{H^{B_1}}\right) = \rho^A \otimes \frac{1}{\dim(H^{B_1})} \sum_{k=1}^{\dim(H^{B_1})} \sigma^B_{\beta_k}$. It is clear that $\frac{1}{\dim(H^{B_1})} \sum_{k=1}^{\dim(H^{B_1})} \sigma^B_{\beta_k} \in \mathcal{B}(H^B)$. To prove (iii)$\Rightarrow$(ii), let $\{|\beta_k\rangle\}_{k=1,\ldots,\dim(H^{B_1})}$ be a normal orthogonal basis of $H^{B_1}$. We can obtain $\forall k \exists \sigma^B_{\beta_k} \forall \rho^A : \varepsilon(\rho^A \otimes |\beta_k\rangle\langle\beta_k|) = \rho^A \otimes \sigma^B_{\beta_k}$ according to (iii) and the lemma 1. Since $\{|\beta_k\rangle\}_{k=1,\ldots,\dim(H^{B_1})}$ is chosen arbitrarily, we have (ii). ∎

Now we define new noiseless subsystems as follows. Let $\varepsilon$ be a quantum channel and decompose a Hilbert space $H = C \oplus C^\perp$ where $C = H^A \otimes H^B$ and $H^B = H^{B_1} \oplus H^{B_2}$. For convenience, let $r \equiv \dim(H^B)$, $r_1 \equiv \dim(H^{B_1})$, and $\{|\beta_j\rangle\}_{j=1,\ldots,r}$ be a normal orthogonal basis of $H^B$. Without loss of generality, $H^{B_1}$ is spanned by $\{|\beta_j\rangle\}_{j=1,\ldots,r_1}$.

**Definition 1.** The subsystem $A$ is called an *ampliate noiseless subsystem* for $(\varepsilon, B_1, B)$ if it satisfies one of (i)-(iii) in the lemma 2.

From the above definition, it is clear that every normal noiseless subsystem can be regarded as a special ampliate noiseless one, that is, the ampliate noiseless subsystem for $(\varepsilon, B_1, B_1)$ is just corresponding to a normal noiseless one.

It is crucial to determine whether $A$ is an ampliate noiseless subsystem for a fixed triple $(\varepsilon, B_1, B)$. One might also ask whether there is a sufficient and necessary condition about this. In this following, we try to answer this question.

We denote $I^{H^A} \otimes |\beta_k\rangle\langle\beta_l|$ by $P_{kl}$ for all $k, l \in \{1, 2, ..., r\}$. Then a map $\Gamma: \mathcal{B}(H) \to \mathcal{B}(H)$ is defined as

$$\Gamma(\rho) = \sum_{k=1}^{r_1} \sum_{l=1}^{r} P_{kl} \rho P_{lk}. \tag{11}$$

Clearly, this map has the following properties: (i) $\Gamma$ is a completely positive and linear map; (ii) $\Gamma(\rho) \propto \sigma^A \otimes I^{H^{B_1}}$; (iii) $\Gamma(\rho^A \otimes \rho^B) \propto \rho^A \otimes I^{H^{B_1}}$. Moreover, we give several notations:

$$\mathcal{P}_B \equiv \sum_{k=1}^{r} P_{kk} = I^{H^A} \otimes I^{H^B}, \quad \mathcal{P}_{B_1} \equiv \sum_{k=1}^{r_1} P_{kk} = I^{H^A} \otimes I^{H^{B_1}}, \text{ and } \mathcal{P}_B^\perp \equiv I^H - \mathcal{P}_B.$$

**Theorem 1.** Given a channel $\varepsilon = \{E_a\}$, $A$ is an ampliate noiseless subsystem for $(\varepsilon, B_1, B)$ if and only if the following two conditions hold:

$$\forall a \forall i, k \in \{1, ..., r_1\} \forall j, l \in \{1, ..., r\}: P_{kl} E_a P_{ij} = \lambda_{aijkl} P_{kj} \tag{12}$$

where $\lambda_{aijkl}$ is some complex number; and

$$\forall a: \mathcal{P}_B^\perp E_a \mathcal{P}_{B_1} = 0. \tag{13}$$

*Proof.* (i) "only if". According to the properties of $\Gamma$ and the lemma 2, we can obtain that

$$(\Gamma \circ \varepsilon \circ \Gamma)(\rho) \propto \Gamma(\rho) \text{ for all } \rho \in \mathcal{B}(H).$$

From linearity of $\Gamma$, the proportionality factor cannot depend on $\rho$. According to the theorem 8.2 in [6], we can obtain that

$$\forall a \forall i, k \in \{1, ..., r_1\} \forall j, l \in \{1, ..., r\}: P_{kl} E_a P_{ij} = \sum_{k'=1}^{r_1} \sum_{l'=1}^{r} w_{aijklk'l'} P_{k'l'}$$

where $\{w_{aijklk'l'}\}$ is some set of complex numbers. Then we have

$$P_{kl} E_a P_{ij} = P_{kk} P_{kl} E_a P_{ij} P_{jj} = P_{kk} \left( \sum_{k'=1}^{r_1} \sum_{l'=1}^{r} w_{aijklk'l'} P_{k'l'} \right) P_{jj} = w_{aijklkl} P_{kj} \equiv \lambda_{aijkl} P_{kj}.$$

It is known that $\varepsilon[\Gamma(\rho)] \in \mathcal{B}(H^A \otimes H^B)$ for all $\rho \in \mathcal{B}(H)$. Since $\mathcal{P}_B^\perp \sigma \mathcal{P}_B^\perp = 0$ for all $\sigma \in \mathcal{B}(H^A \otimes H^B)$, we can obtain $\mathcal{P}_B^\perp \varepsilon[\Gamma(\rho)] \mathcal{P}_B^\perp = 0$ for all $\rho \in \mathcal{B}(H)$. By the definition of $\Gamma$, $\mathcal{P}_B^\perp \varepsilon[\Gamma(\rho)] \mathcal{P}_B^\perp = \sum_a \sum_{k=1}^{r_1} \sum_{l=1}^{r} \mathcal{P}_B^\perp E_a P_{kl} \rho P_{lk} E_a^\dagger \mathcal{P}_B^\perp$. According to the theorem 8.2 in [6], we can obtain $\mathcal{P}_B^\perp E_a P_{kl} = 0$ for all $k \in \{1,2,...,r_1\}$, $l \in \{1,2,...,r\}$, and $a$. Thus $\mathcal{P}_B^\perp E_a \mathcal{P}_{B_1} = 0$.

(ii) "if". For all $\rho \in \mathcal{B}(H^A \otimes H^{B_1})$, we can get

$$\begin{aligned}
\varepsilon(\rho) &= (\mathcal{P}_B + \mathcal{P}_B^\perp) \varepsilon(\rho)(\mathcal{P}_B + \mathcal{P}_B^\perp) \\
&= \mathcal{P}_B \varepsilon(\rho) \mathcal{P}_B + \mathcal{P}_B \varepsilon(\rho) \mathcal{P}_B^\perp + \mathcal{P}_B^\perp \varepsilon(\rho) \mathcal{P}_B + \mathcal{P}_B^\perp \varepsilon(\rho) \mathcal{P}_B^\perp \\
&= \mathcal{P}_B \varepsilon(\rho) \mathcal{P}_B + \mathcal{P}_B \varepsilon(\mathcal{P}_{B_1} \rho \mathcal{P}_{B_1}) \mathcal{P}_B^\perp + \mathcal{P}_B^\perp \varepsilon(\mathcal{P}_{B_1} \rho \mathcal{P}_{B_1}) \mathcal{P}_B + \mathcal{P}_B^\perp \varepsilon(\mathcal{P}_{B_1} \rho \mathcal{P}_{B_1}) \mathcal{P}_B^\perp \\
&= \mathcal{P}_B \varepsilon(\rho) \mathcal{P}_B
\end{aligned}$$

where the third identity follows from (13). Thus for all $\rho^A$ and $\rho^{B_1}$, we have

$$\varepsilon(\rho^A \otimes \rho^{B_1}) = \mathcal{P}_B \varepsilon\left[\mathcal{P}_{B_1}(\rho^A \otimes \rho^{B_1}) \mathcal{P}_{B_1}\right] \mathcal{P}_B$$

$$= \sum_a \sum_{k,l=1}^{r} \sum_{i,j=1}^{r_1} P_{kk} E_a P_{ii} (\rho^A \otimes \rho^{B_1}) P_{jj} E_a^\dagger P_{ll}$$

$$\overset{\substack{s,v \in \{1,...,r_1\} \\ t,u \in \{1,...,r\}}}{=} \sum_a \sum_{k,l=1}^{r} \sum_{i,j=1}^{r_1} P_{ks} P_{sk} E_a P_{it} P_{ti} (\rho^A \otimes \rho^{B_1}) P_{ju} P_{uj} E_a^\dagger P_{lv} P_{vl}$$

$$= \sum_a \sum_{k,l=1}^{r} \sum_{i,j=1}^{r_1} P_{ks} \lambda_{aitsk} P_{st} P_{ti} (\rho^A \otimes \rho^{B_1}) P_{ju} \overline{\lambda}_{ajuvl} P_{uv} P_{vl}$$

$$= \rho^A \otimes \left( \sum_a \sum_{k,l=1}^{r} \sum_{i,j=1}^{r_1} \lambda_{aki} \overline{\lambda}_{alj} |\beta_k\rangle\langle\beta_i| \rho^{B_1} |\beta_j\rangle\langle\beta_l| \right)$$

Clearly, $\sum_a \sum_{k,l=1}^{r} \sum_{i,j=1}^{r_1} \lambda_{aki} \overline{\lambda}_{alj} |\beta_k\rangle\langle\beta_i| \rho^{B_1} |\beta_j\rangle\langle\beta_l| \in \mathcal{B}(H^B)$. ∎

The above theorem shows that normal noiseless subsystems are not equivalent to ampliate noiseless ones; and the latter is more generalized than the former. In fact, according the theorem 1 in [14], if the subsystem $A$ is a normal noiseless one with its corresponding noisy subsystem $B_1$ then the condition $\forall a : \mathcal{P}_{B_1}^\perp E_a \mathcal{P}_{B_1} = 0$ should be satisfied. Clearly, this condition is stricter than (13) when $\dim(H^{B_1}) < \dim(H^B)$. Thus amplite noiseless subsystem method is new one for quantum error correction. Based on amplite noiseless subsystems, we will introduce a more unified framework for QEC than OQEC as follows.

## 3. Generalized OQEC (GOQEC) Approach

Given a fixed decomposition $H = C \oplus C^\perp$ where $C = H^A \otimes H^B$ and $H^B = H^{B_1} \oplus H^{B_2}$, let us to define a density operator set $\mu = \{\rho^A \otimes \rho^{B_1} \mid \rho^A \in \mathcal{B}(H^A), \rho^{B_1} \in \mathcal{B}(H^{B_1})\}$. The approach for GOQEC consists of a quadruple $(\mathcal{R}, \varepsilon, \mu, B)$ where the recovery $\mathcal{R}$ and error $\varepsilon$ are quantum channels on $\mathcal{B}(H)$. In particular, $(\mathcal{R}, \varepsilon, \mu, B_1)$ is just equivalent to $(\mathcal{R}, \varepsilon, \mu)$ [14] in OQEC. Note that $(\mathcal{R}, \varepsilon, \mu)$ is corresponding to the decomposition $H = C' \oplus C'^\perp$ where $C' = H^A \otimes H^{B_1}$. Then we can give the following definition by using the definition 1.

**Definition 2.** Given a quadruple $(\mathcal{R}, \varepsilon, \mu, B)$, we say that it is *correctable* if the following condition holds: For some density operator $\rho \in \mathcal{B}(H^B)$, we have

$$\forall \rho^A : (\mathcal{R} \circ \varepsilon)\left(\rho^A \otimes \frac{1}{r_1} I^{H^{B_1}}\right) = \rho^A \otimes \rho. \tag{14}$$

From the above definition, it is trivial that if $(\mathcal{R}, \varepsilon, \mu)$ in OQEC is correctable then $(\mathcal{R}, \varepsilon, \mu, B)$ is also correctable. Clearly, the converse proposition does not hold since there are some amplite noiseless subsystems which are not "normal".

**Definition 3.** Given a channel $\varepsilon$ and a fixed decomposition $H = C' \oplus C'^\perp$ where $C' = H^A \otimes H^{B_1}$, we say that $\mu$ is *correctable* for $\varepsilon$ in GOQEC if there exists a quantum recovery operation $\mathcal{R}$ and a decomposition $H = C \oplus C^\perp$ with $C = H^A \otimes H^B$ and $H^B = H^{B_1} \oplus H^{B_2}$ such that $(\mathcal{R}, \varepsilon, \mu, B)$ is correctable. Note that $\mu$ is defined as above.

From the above definition, it is trivial that if $\mu$ is correctable [14] for $\varepsilon$ in OQEC then it is also correctable for $\varepsilon$ in GOQEC. Now ask whether the converse proposition is true. One might answer "No" immediately since normal noiseless subsystems are not equivalent amplite noiseless ones. But it is very interesting that our answer is "Yes". To see this, we first prove a sufficient and necessary condition for quantum error correction by using GOQEC approach as follows.

**Theorem 2.** Let $\varepsilon = \{E_a\}$ be a channel and Decompose a Hilbert space $H = C' \oplus C'^\perp$ where $C' = H^A \otimes H^{B_1}$. $\mu$ is correctable for $\varepsilon$ in GOQEC if and only if there exists a

complex number set $\{\lambda_{abkl}\}$ such that

$$\forall a,b \forall k,l \in \{1,...,r_1\}: P_{kk}E_a^\dagger E_b P_{ll} = \lambda_{abkl} P_{kl}. \tag{15}$$

*Proof.* (i) "only if". According to the definitions 2 and 3, there exists a recovery operation $\mathcal{R} = \{R_c\}$ and a decomposition $H = C \oplus C^\perp$ with $C = H^A \otimes H^B$ and $H^B = H^{B_1} \oplus H^{B_2}$ such that $\forall \rho^A : (\mathcal{R} \circ \varepsilon)\left(\rho^A \otimes \frac{1}{r_1} I^{H^{B_1}}\right) = \rho^A \otimes \rho$ for some density operator $\rho \in \mathcal{B}(H^B)$.

Thus we can obtain that

$$P_{kk}E_a^\dagger E_b P_{ll} = P_{kk} E_a^\dagger \sum_c R_c^\dagger (\mathcal{P}_B + \mathcal{P}_B^\perp) R_c E_b P_{ll}$$

$$= P_{kk} E_a^\dagger \sum_c R_c^\dagger P_B R_c E_b P_{ll}$$

$$\overset{j\in\{1,...,r_1\}}{=} \sum_c \sum_{i=1}^r P_{kk} E_a^\dagger R_c^\dagger P_{ij} P_{ji} R_c E_b P_{ll}$$

$$= \sum_c \sum_{i=1}^r \overline{\lambda}_{ackij} \lambda_{cbjil} P_{kj} P_{jl}$$

$$= \left(\sum_c \sum_{t=1}^r \overline{\lambda}_{ackij} \lambda_{cbjil}\right) P_{kl} \equiv w_{abkl} P_{kl}$$

where the second and fourth identities follow from the theorem 1.

(ii) "if". According to the theorem 2 in [14], we can obtain that $\mu$ is correctable for $\varepsilon$ in OQEC. Thus $\mu$ is also correctable for $\varepsilon$ in GOQEC. ∎

Clearly, the condition (15) in the above theorem is the same as one of the theorem 2 in [14]. Thus, for a fixed decomposition $H = C' \oplus C'^\perp$ where $C' = H^A \otimes H^{B_1}$, $\mu$ is correctable for $\varepsilon$ in OQEC if and only if $\mu$ is correctable for $\varepsilon$ in GOQEC. This means that GOQEC model does not bring new forms of codes. However, GOQEC provides more recovery operations than OQEC. This is very important for some experiments of QEC because GOQEC possibly provides simpler operations to detect and recover errors. To see this, we first show how OQEC and GOQEC codes can be transformed each other.

**Theorem 3.** Suppose that $H = C \oplus C^\perp$ with $C = H^A \otimes H^B$ and $H^B = H^{B_1} \oplus H^{B_2}$. Let $\mathcal{R}, \varepsilon$ be two quantum channels on $\mathcal{B}(H)$, $\{|\alpha_k\rangle\}$ be a normal orthogonal basis of $H^{C^\perp}$ and $\mu$ be defined as above. Then

(i) If $(\mathcal{R}, \varepsilon, \mu)$ is correctable then $(\eta_1 \otimes I^{C^\perp} \circ \mathcal{R}, \varepsilon, \mu, B)$ is correctable, where

$$\eta_1(\bullet) = \frac{1}{r}\sum_{k,l=1}^{r} P_{kl}(\bullet)P_{lk} + \sum_{k=1}^{\dim\left(H^{C^\perp}\right)} |\alpha_k\rangle\langle\alpha_k|(\bullet)|\alpha_k\rangle\langle\alpha_k|;$$

(ii) If $(\mathcal{R},\varepsilon,\mu,B)$ is correctable then $(\eta_2 \circ \mathcal{R}, \varepsilon, \mu)$ is correctable, where

$$\eta_2(\bullet) = \frac{1}{r_1}\sum_{k=1}^{r_1}\sum_{l=1}^{r} P_{kl}(\bullet)P_{lk} + \sum_{k=1}^{\dim\left(H^{C^\perp}\right)} |\alpha_k\rangle\langle\alpha_k|(\bullet)|\alpha_k\rangle\langle\alpha_k|.$$

*Proof.* (i) It is clear that $\eta_1$ is also a completely positive and trace-preserving linear map on $\mathcal{B}(H)$. For all $\rho^A$, we can obtain that

$$(\eta_1 \circ \mathcal{R} \circ \varepsilon)\left(\rho^A \otimes \frac{1}{r_1} I^{H^{B_1}}\right) \stackrel{\exists \sigma^{B_1}}{=} \eta_1(\rho^A \otimes \sigma^{B_1})$$

$$= \frac{1}{r}\sum_{k,l=1}^{r} P_{kl}(\rho^A \otimes \sigma^{B_1})P_{lk}$$

$$= \rho^A \otimes \frac{1}{r}\sum_{k,l=1}^{r} |\beta_k\rangle\langle\beta_l|\sigma^{B_1}|\beta_l\rangle\langle\beta_k|$$

Clearly, $\frac{1}{r}\sum_{k,l=1}^{r}|\beta_k\rangle\langle\beta_l|\sigma^{B_1}|\beta_l\rangle\langle\beta_k| \in \mathcal{B}(H^B)$.

(ii) It is similar to (i). ∎

Recall the example in the introduction. We first use the GOQEC approach to correct errors $\varepsilon = \{E_0, E_1, E_2\}$ as follows. We decompose the Hilbert space $\mathbb{C}^4 = H^A \otimes H^B$ with the computational basis, so that $H^A = \mathbb{C}^2$, $H^B = H^{B_1} \oplus H^{B_2}$, $H^{B_1} = \sup(|0\rangle\langle 0|)$ and $H^{B_2} = \sup(|1\rangle\langle 1|)$. Then $\mu = \{\rho^A \otimes |0\rangle\langle 0| \mid \rho^A \in \mathcal{B}(H^A)\}$. Clearly, $(I_2 \otimes I_2, \varepsilon, \mu, B)$ is correctable, which means that no recovery operation is done. According to the above theorem, we can also use OQEC approach to correct errors. Although $(\eta, \varepsilon, \mu)$ where

$$\eta(\bullet) = \sum_{k=0}^{1}(I_2 \otimes |0\rangle\langle k|)(\bullet)(I_2 \otimes |k\rangle\langle 0|)$$ is also correctable, the operation $\eta$ must be done.

**4. Conclusions**

We present a more generalized notion for noiseless subsystems, called ampliate noiseless subsystems. The quantum information is encoded in $A$ of $H^A \otimes H^{B_1}$. The normal noiseless subsystem model in [14, 15] admits the state in noisy subsystem $B_1$ to be mapped to a state in

$B_1$ under a channel $\varepsilon$ while in our notion $\varepsilon$ could in principle map the state in $B_1$ to a state in any subsystem $B$ (including $\dim(H^{B_1}) < \dim(H^B)$) satisfying $H^A \otimes H^B$. Base on ampliate noiseless subsystems, we introduce a more unified approach to quantum error correction, so-called GOQEC. We obtain an important consequence: the existence of OQEC codes is equivalent to one of GOQEC codes. While GOQEC model does not lead to new families of codes, it does allow for new error correction procedures, possibly enriching the fault tolerance quantum computing theory. For example, it is possible for some experiments of QEC that GOQEC codes are easier to detect and recover errors than OQEC ones because the former provides more recovery operations than the latter. This implies that we might construct simpler fault-tolerant gates by means of GOQEC.

This work is supported by the Chinese National Program on Key Basic Research Project (973 Program, Grant Nos. 2014CB744605 and 2013CB329304), the Natural Science Foundation of China (Grant Nos. 61170178, 61272254 and 61272265).